\documentclass[twocolumn,prl,showpacs,preprintnumbers,amsmath,amssymb]{revtex4}

\usepackage{graphicx}   
\usepackage{dcolumn}   
\usepackage{bm}            
\usepackage{textcomp}
\usepackage{epsfig}
\usepackage{psfig}
{\rm }

\begin{document}

\preprint{APS}

\title{Pattern formation driven by  nematic ordering of assembling biopolymers}

\author{Falko Ziebert}
\author{Walter Zimmermann}
\affiliation{
Theoretische Physik, Universit\"at  des Saarlandes, D-66041 Saarbr\"ucken, Germany
}

\date{\today}

\begin{abstract}
The biopolymers actin and microtubules are often in an
ongoing  assembling/disassembling state far from thermal
equilibrium. Above a critical density this leads to spatially
periodic patterns, as shown by a scaling argument and in terms
of a phenomenological continuum model, that meets   also
 Onsager's statistical theory of the nematic--to--isotropic transition
in the absence of reaction kinetics.
  This pattern forming process
depends   much on nonlinear effects and a
common linear stability analysis of the isotropic  distribution of the filaments
is often misleading. The wave number  of the pattern decreases
with the assembling/disassembling rate and there
is an uncommon discontinuous
transition between the nematic and the periodic state.
\end{abstract}

\maketitle

Ongoing polymerization and depolymerization
of actin and  microtubule
filaments are prominent   examples for
dissipative non--equilibrium phenomena  in living cells
\cite{Alberts:2001}, which are
important for many different purposes, such as
 the cell motility and division or morphogenesis.
Both substances show  also  an
 inherent propensity to  pattern formation  and active phenomena
\cite{Hill:87.1,Mandelkow:89.1,Tabony:90.1,Leibler:97.1,Ott:1997.1,Kaes:2002.1,Kaes:2003.101}.
Like the famous example for rod--like particles,  the Tobacco Mosaic Virus (TMV)
\cite{Bawden:1936.1}, also actin and microtubule filaments
may undergo with increasing density
  a well known transition to an orientational order
  \cite{Hitt:90.1,Suzuki:91.1,Kaes:96.1},
the so--called  nematic order \cite{Onsager:49.1,deGennes:93}.
By Onsager's seminal work \cite{Onsager:49.1,Straley:1973.1,deGennes:93,Odijk:86.1}
this  transition has been traced back
to excluded volume interactions between the filaments.
This statistical theory is valid for long filaments
of fixed shape and infinite lifetime $\tau$,
 and it   predicts  near the transition
also  a phase separation into  domains of
isotropically
oriented rods at low density and nematic domains of
higher rod--density, which has
also  been observed for actin with  an almost  vanishing
kinetics \cite{Kaes:96.1}. For a finite lifetime
of actin and microtubule filaments, Onsager's
statistical theory for the nematic order does not apply.
 Moreover, a finite $\tau$ limits the diffusive
transport distance and the coarsening during the
 phase separation close to the orientational transition
to  a length scale of about $ l_{D}=\sqrt{ D_\rho\tau}$,
 with the filament diffusion constant $D_\rho$.
According to our estimate we expect  kinetically induced
periodic patterns with a wavelength in the order of
 10  $\mu m$. This is supported by current
experiments \cite{Kaes:2003.101}.

The effect of   reaction kinetics on a phase separation
has  been investigated for a  chemical   and a biophysical
example in Refs. \cite{Glotzer:95.1,Zimmermann:95.2}.
In both cases the transition to periodic patterns
is supercritical  and its onset  as well as wavelength  follows already
from a  linear stability analysis of the
respective  homogeneous basic state.
Instead of these  two competing states, near
 the orientational transition of filaments  one has  three different competing
states, the spatially homogeneous
isotropic state, the spatially homogeneous
nematic one and the spatially inhomogeneous alternation
between  the isotropic and the nematic order.
Here, the growth rate of perturbations of the basic state takes its  maximum also at a finite wavenumber
\cite{Doi:88.1,Liverpool:2003.2}, but this is not
 sufficient for a prediction of spatially inhomogeneous
nonlinear states  above
the orientational transition.
Instead of an inhomogeneous state, as suggested
by the linear perturbation analysis,
 in the nonlinear regime one has an exchange of stability
and the spatially homogeneous nematic state is often preferred,
as described in this work. Therefore the  bifurcation picture
has to be explored by a nonlinear analysis, whereby
 an uncommon coarsening behavior at the nonlinear
nematic--to--periodic transition has been found.

The generic  scenario near the ordering transition
is  described  in terms of a phenomenological model,
 which is introduced and analyzed
at  first  without the reaction kinetics of the filaments.
 It is extended in the second part
by the  essential reaction steps as motivated by
actin and microtubule polymerization.

\begin {figure}[htp]
 \epsfig{figure=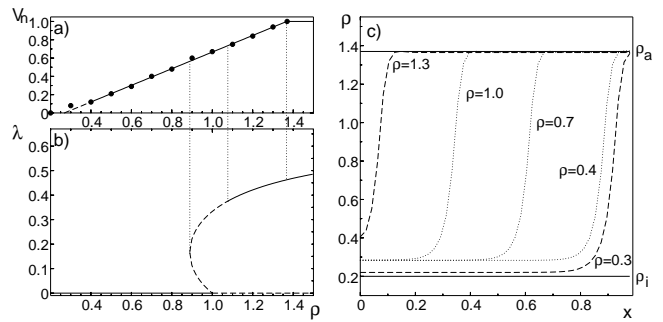,width=8.6cm,angle=00}
\caption{Part b) shows the stable (solid) and the
unstable branch (dashed) of the
nematic order parameter $\lambda_{\pm}$
as a function of the rod density $\rho$, cf. Eq.~(\ref{lambnem}).
 Part c) shows  stable kink solutions
of Eqs.~(\ref{rl}) interpolating
between  the nematic ($\rho=\rho_a$)
and  the isotropic range ($\rho=\rho_i$).
The nematic volume fraction $V_n$ as a function of $\rho$ is given in a).
Parameters of the model:
  $D_{r}=0.1,~D_{\rho}=D_{\lambda}=0.3,~a_{\rho}=0.25,~a_{\lambda}=2.0,
\delta_{\rho}=\delta_{\lambda}=0.1$. For this set, the whole  system
is in the homogeneous nematic state for  $\rho_0>1.367$  (i.e $V_{n}=1$).
}
\label {fig_bifurc}
\vspace{-2mm}
\end {figure}

{\it Model without reaction kinetics. --}
%
In lyotropic liquid crystals
the nematic order is forced beyond a critical rod
density  $\tilde \rho_c$ by  excluded volume interactions \cite{Onsager:49.1}
and the resulting
  local mean orientation of
rod--like particles is described by the so--called
director ${\bf n}( {\bf r})$ (with  ${\bf n}=-{\bf n}$)
 \cite{deGennes:93}.
Assuming  rods of a single length and
a uniform  ${\bf n}$,
 then the largest eigenvalue
$\tilde \lambda$ of the nematic order
 parameter tensor is sufficient for a description of the strength of the
orientational order.
  $ \tilde \lambda$ varies in the range
$[\frac{1}{3}, 1]$, with
$\tilde \lambda=\frac{1}{3}$ in the isotropic state  and  $\tilde \lambda=1$
for  a uniform  rod orientation \cite{deGennes:93,Dhont:2001.1}.
It is convenient  to use  the difference
$\lambda :=\tilde \lambda - \frac{1}{3}$
 and
the dimensionless rod density  $\rho \propto \tilde \rho V_E$,
with $V_E=2bL^2$ the excluded volume for rods of length
$L$ and diameter $b$.
For a constant  orientation ${\bf n}$, as we assume in
this work,
spatial variations of  $\rho$ and $\lambda$  also include  spatially alternating   isotropic and nematic ranges.
 Since for  long rods the preferred director orientation
is parallel to the isotropic--nematic interface \cite{Liu:70.1}, we assume
spatial variations in the  direction  perpendicular to  ${\bf n}$, which we call the   $x$--direction.
For this context we choose the  phenomenological
model for the conserved density $\rho(x,t)$ and for the unconserved
field $\lambda(x,t)$
\begin{subequations}\label{rl}
\begin{eqnarray}\label{r1}
\partial_{t}\rho&=&D_{\rho}\partial_{x}^{2}\left[-\lambda\rho-\delta_{\rho}\partial_{x}^{2}\rho+a_{\rho}\rho^{3}\right]~,\\
\label{l1}
\partial_{t}\lambda&=&-D_{r}\left[(1-\rho)\lambda-\frac{3}{2}\rho\lambda^{2}+\frac{9}{2}\rho\lambda^{3}\right]
\nonumber\\
& & +D_{\lambda}\partial_{x}^{2}\left[(1-\rho)\lambda-\delta_{\lambda}\partial_{x}^{2}\lambda+a_{\lambda}\lambda^{3}\right].
\end{eqnarray}
\end{subequations}
In Eq.~(\ref{l1}), the part  without spatial derivatives
 follows   from the
Smoluchowski equation for rigid rods by a moment approximation
\cite{Dhont:2001.1,Ziebert:2003.02} and  the factor
  $\rho$ in front of the
nonlinear terms  reflects the excluded volume interaction.
It determines also the homogeneous solutions
\begin{equation}
\label{lambnem}
\rho=\rho_0=const. \,, \quad \lambda_{0}=0, \quad
\lambda_{\pm}=\frac{1}{6}\pm\frac{1}{6}\sqrt{9-\frac{8}{\rho_0}~}~.
\end{equation}
$\lambda_0=0$ corresponds
to the isotropic rod distribution that  becomes
linearly unstable with respect to
nematic  fluctuations
 beyond the critical density $\rho > \rho_c=1$,
where they  grow up to the homogeneously stable
upper branch
$\lambda_{+}$ of the
spatially uniform  nematic order in Fig.~\ref{fig_bifurc}b).
Since the isotropic-nematic transition
is of first order, both states
coexist in a range $\frac{8}{9} \le \rho \le 1$.

Eq.~(\ref{r1}) is of the Cahn-Hilliard type \cite{Cahn:58.1}.
Expressing its right hand side by a    divergence  of the  current density
$j_\rho(x)=-D_{\rho}\partial_x \mu(x)$ with $\mu(x)=
  -\lambda \rho- \delta_\rho \partial_x^2
\rho +a_\rho \rho^3$,   it takes the form of
a  conservation law for the rod--like particles.
 The first (nonlinear) term in Eq.~(\ref{r1}),
i.e. $-D_\rho \partial_x^2 (\lambda \rho )$,  destabilizes the
spatially homogeneous particle density
for any finite value of $\lambda$ ($\lambda$ is always positive) and  mimics
therefore Onsager´s prediction  \cite{Onsager:49.1,Odijk:86.1}
that the free energy can be reduced, by separating the system
  into  ranges of low  rod density $\rho_i$
(isotropic) and
high  density $\rho_a$ (nematic).
The second term describes an
isotropic-nematic interface energy and the third term limits
the modulation amplitudes of the  density (see also below).

%
%
%
Theories involving the distribution function predict
beyond the critical  $\rho_c=1$
 an instability
of the isotropic distribution
 against   inhomogeneous order parameter fluctuations
\cite{Doi:88.1,Ziebert:2003.02}.
In  Eq.~(\ref{l1}) this
is taken into account  by  $\partial_x^2 ((1-\rho)\lambda)$
and the last two terms
limit the wavenumber and the
 amplitude of the nonlinear modulations of $\lambda$.
For  intermediate values of   $\rho$,
Eqs.~(\ref{rl})
have stationary kink solutions as shown in  Fig.~\ref{fig_bifurc}c).
 The densities $\rho_i$ and $\rho_a$, in the
isotropic and nematic range respectively,  are
determined by the two   coefficients
 $a_\rho$ and $a_\lambda$, as described in the
following.
For stationary  kinks as in Fig.~\ref{fig_bifurc}c) the
 particle transport vanishes $j_\rho(x)=0$ and
$\mu(x)=\mu_i=\mu_a $ is constant.
Sufficiently far away from the kink   $\rho_a$ and $\rho_i$ are
 constant too and one obtains
the equation
\begin{eqnarray}
\label{muai}
a_\rho  \rho_i^3 = -\lambda_a \rho_a + a_\rho \rho_a^3 \,,
\end{eqnarray}
where  $\lambda_{a}=\lambda(\rho_a)= ( 1 +\sqrt{9-8/\rho_a\,}~)/6 $.
The rotational term in Eq.~(\ref{l1}) vanishes in the nematic range as well as
trivially in the isotropic range. Defining $j_{\lambda}(x)=-D_{\lambda}\partial_x \nu(x)$, since
in the isotropic range $\lambda_{i}$ is zero, $\nu_{i}$ is zero as well.
To prevent a current through the interface, the total current, i.e. just
$j_{\lambda}$ in the nematic region, has to vanish
and it follows $\nu_{a}=\nu_{i}=0$ leading to
\begin{eqnarray}
 (1-  \rho_a)\lambda_a + a_\lambda \lambda_a^3 =0\,\,.
\end{eqnarray}
As $\lambda_a$ is known, from this equation  the
 anisotropic density $\rho_a$ follows as a function of $a_\lambda$
(or vice versa) and $\rho_i$  is given with
$\rho_a$  via Eq.~(\ref{muai})
as a function of $a_\rho$ (or vice versa). Therefore,
 the  two densities $\rho_i$ and $\rho_a$ may also be
considered as input parameters that are
 obtained from different approaches as
for instance from Refs.~\cite{Onsager:49.1,Kayser:1977.1}
or possibly from  experiments.
 Since $\rho_i$ and   $\rho_a$
do not depend on the system size $L$, the kink position
changes with the   mean density $\rho_0$  as shown in Fig.~\ref{fig_bifurc}, where the
nematic volume fraction $V_n$ is given  in terms  of   $\rho_0$ .

\begin {figure}[htp]
\epsfig{figure=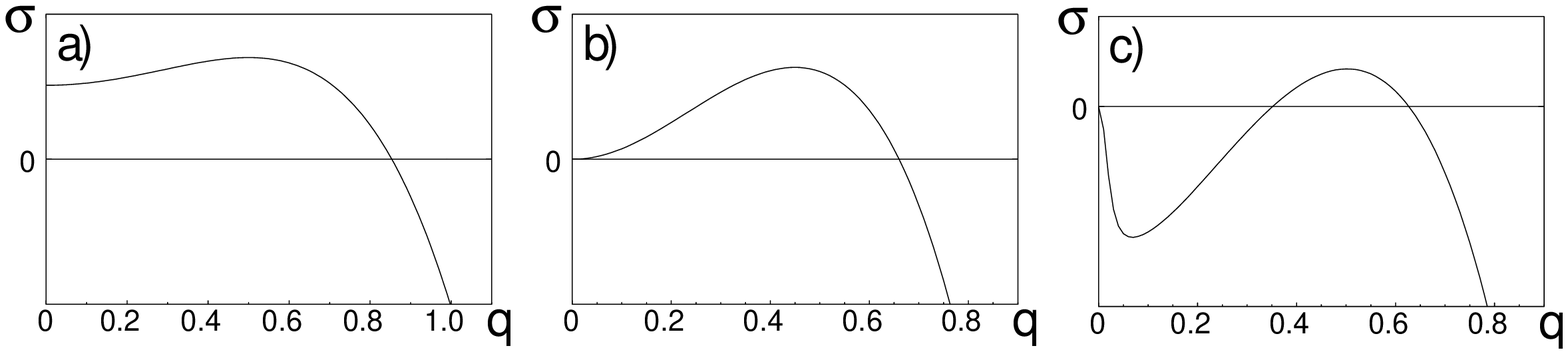,width=8.6cm,angle=00}
\caption{In part  a)  $\sigma(q)$ is shown for periodic perturbations
of
the isotropic state with  $\rho_0 >\rho_c$ and in  part b)
of
the nematic state without  kinetics  and at the
unstable branch in   Fig.~\ref{fig_bifurc}b). In part c)  $\sigma(q)$
is given for perturbations of
 the nematic state in the case  with reaction kinetics  and
at the unstable branch  in
 Fig.~\ref{fig_per}d).
 In part a) and b) the parameters are
 $\rho_0=1.05$ and  $\Sigma=0$  and in c)
$\Sigma=0.003$ and $s=0.01\Sigma$. The other parameters are as in Fig.~\ref{fig_bifurc}.
}
\label {fig_disprel}
\end {figure}

{\it Linear stability analysis. --}
For  $\rho>\rho_c$ a linear stability analysis of the homogeneous isotropic state,
cf. $\lambda=0$ and  $\rho=\rho_0$, with respect to
small periodic  perturbations
$\lambda_1, ~\rho_1 \propto \exp( \sigma t \pm iqx) $  gives the
wave number
 dependence of the growth rate $\sigma(q)$ as shown for a set of parameters
in Fig.~\ref{fig_disprel}a). This shape of $\sigma(q)$ with  a positive
 value at $q=0$ and a  maximum
at a finite value of  $q$ is typical for the unstable isotropic state and
 is in agreement with similar results for microscopic models \cite{Doi:88.1}. A  linear stability
analysis  of the homogeneous nematic state  $\lambda_{\pm}$
in terms of  microscopic models is rather involved
\cite{Ziebert:2003.02}. For
our phenomenological model, however,  the determination of
$\sigma(q)$ is a straightforward task and its typical shape
at the  unstable nematic branch is shown in Fig.~\ref{fig_disprel}b).
Along the dashed part of the curve in
 Fig.~\ref{fig_bifurc}b) the homogeneous nematic state
is linear unstable.  Between
our  result and the linear stability described in ~\cite{Doi:88.1} there is
a major difference.  In both cases
  $\sigma(q)$  for the isotropic state
 takes its maximum at a
finite value of    $q$  and  has positive
values for any  $ \rho>\rho_c$. This is somewhat
 in contradiction
to Onsager's statistical
theory, where inhomogeneous states (via phase separation) are   only
energetically preferred
for a rod density below a maximum value $ \rho < \rho_a$. In
our model  nonlinear effects stabilize the uniform
nematic state  for $\rho>\rho_a$  and it is unstable only along
 the dashed line
 in Fig.~\ref{fig_bifurc}b).
 Simulations of Eqs.~(\ref{rl}) confirm
 that inhomogeneous solutions $\lambda(x)$ and $\rho(x)$ only occur
 for  a mean density  $\rho_0$ smaller than $\rho_a$.
%
%
%

{\it Reaction kinetics drives  pattern. --}
%
%
In cells and in vitro actin and microtubule filaments
are usually  out of  equilibrium and, due to an
ongoing  assembly/disassembly reaction,
filaments have
 a finite lifetime  $\tau$.
This  reaction kinetics  leads to
a stationary length distribution
 of the filaments  \cite{Kaes:96.1} or even to
oscillatory polymerization  \cite{Hill:87.1,Mandelkow:94.1,Hammele:2003.1}.
During the phase separation at  the isotropic--to--nematic
transition, filaments are transported, but
only over a lifetime--dependent  distance
of about  $l_{D}=\sqrt{D_{\rho}\tau}$. Since the  lifetime of filaments
 is a constant,   much more
subunits are released in the nematic range
 with a   high density $\rho_a$  than in the isotropic
  range with a low  density $\rho_i$. However, due to
a much larger diffusion constant, the subunits are redistributed
 quickly, leading to a nearly
   homogeneous subunit density $m(x)$. Thus the number of nucleated
filaments
per unit time, which depends on $m(x)$, is weakly varying too.
By this qualitative
reasoning one expects a steady net transport of subunits from the
nematic to the isotropic range and in the opposite direction a transport of
 filaments, whereby the latter one  is limited to distances
of the order of  $l_{D}$ or smaller. This length restriction  causes, instead of a
large scale  phase separation, a spatially periodic pattern
with a   wavelength in the order of  $l_{D}$.

Along this qualitative reasoning
 the complexity of the biochemical
reaction steps,  involved during the assembly/disassembly  of
actin or microtubules, is   not crucial
for this wavelength limitation.
For instance,  actin and
microtubules are usually met with a
 polydisperse length distribution.
 Since the
slowest kinetic step and the small diffusion constant of the long filaments
 will govern the limitation, we discard the polydispersity and
assume
for the sake of simplicity that all filaments are of the same length.
With a decay  $\Sigma=\tau^{-1}$ and a nucleation rate $s$ of the
filaments and  a diffusion constant $D_m$
of the subunits
  one ends up finally with the three equations
\begin{subequations}\label{rl2}
\begin{eqnarray}\label{r2}
\partial_{t}\rho&=&D_{\rho}\partial_{x}^{2}\left[-\lambda\rho-\delta_{\rho}\partial_{x}^{2}\rho+a_{\rho}\rho^{3}\right]+sm-\Sigma\rho,\\
\label{l2}
\partial_{t}\lambda&=&-D_{r}\left[(1-\rho)\lambda-\frac{3}{2}\rho\lambda^{2}+\frac{9}{2}\rho\lambda^{3}\right]
\nonumber\\
& & \hspace{-3mm}
+D_{\lambda}\partial_{x}^{2}\left[(1-\rho)\lambda-\delta_{\lambda}\partial_{x}^{2}\lambda+a_{\lambda}\lambda^{3}\right]-\Sigma\lambda~,\\
\label{m2}
\partial_{t}m&=&D_{m}\partial_{x}^{2}m-\gamma sm+\gamma\Sigma\rho~.
\end{eqnarray}
\end{subequations}
The constant $\gamma$ is a measure for the
number of subunits per rod--like particle. In the  nematic range
oriented filaments are lost, but
new ones are nucleated everywhere with
an arbitrary orientation which have to  relax to
the local mean orientation by rotational diffusion.
Accordingly, there is only a decay term  in Eq.~(\ref{l2})
which  can be justified also microscopically
\cite{Ziebert:2003.02}.  This reaction kinetically caused
 partial loss of the orientational order leads
to a higher  critical density for the
 isotropic--to--nematic transition  $\rho_c = 1+ \Sigma/D_r$.
%
\begin{figure}
\epsfig{figure=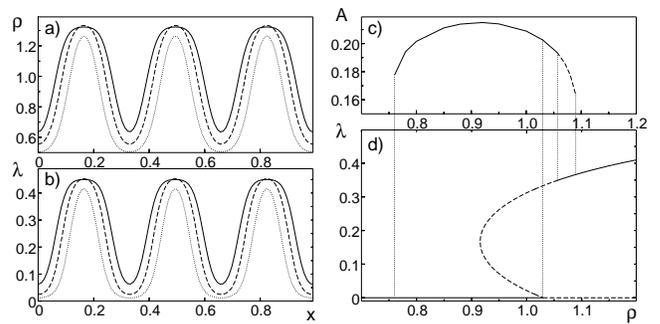,width=8.6cm,angle=00}
\caption{Part a) and b) show periodic
solutions  $\rho(x)$ and $\lambda(x)$  of  Eqs.~(\ref{rl2}) for
 $\rho_0=0.8$ (dotted),  $\rho_0=0.95$ (dashed)
and $\rho_0=1.056$ (solid), respectively.
Part d) displays the  stable homogeneous nematic branch (solid)
and the unstable ones (dashed).  In part c)
 the existence range of
the stable (solid) and unstable (dashed) periodic patterns
with the  modulation amplitude $A$ of  $\lambda(x)$ are given.
Parameters are as in Fig. \ref{fig_bifurc} with $D_{m}=10,~\gamma=100,~\Sigma=0.003,~s=0.01\Sigma$.
}
\label{fig_per}
\vspace{-2mm}
\end{figure}
%
%
In the presence of reactive steps the rates $s$ and $\Sigma$
determine the mean rod density in terms of  the  monomer density
$\rho =s \Sigma^{-1}m$. The  spatially homogeneous
solutions of Eqs.(\ref{rl2}) are
\begin{equation}
\label{lambnem2}
\rho=\rho_0 \,,~~ \lambda_{0}=0, ~~
\lambda_{\pm}=\frac{1}{6}\pm\frac{1}{6}
\sqrt{9-\frac{8}{\rho_0}\left(1+\frac{\Sigma}{D_{r}}\right)}~,
\end{equation}
where  $\lambda_{\pm}(\rho)$
are depicted as a function of the mean density $\rho$
in Fig.~\ref{fig_per}d). In a certain parameter range $\lambda_+$
may become unstable  with respect to periodic perturbations
$\lambda_1, ~\rho_1 \propto \exp( \sigma t \pm iqx) $ and the
 wave number dependence of the growth rate  $\sigma(q)$
 has  a typical shape as  in Fig.~\ref{fig_disprel}c).
 Compared to the case without kinetics
as in  Fig.~\ref{fig_disprel}b),  long wavelength
 perturbations are now suppressed  and only
perturbations with a finite wave number grow.
 The parameter range
of the unstable homogeneous $\lambda_+$ branch is
 indicated  by   the dashed
and the stable one by the solid line in Fig.~\ref{fig_per}d).
The unstable range of the nematic branch decreases
with increasing coefficients $\delta_\rho$ and $\delta_\lambda$
and with increasing  and moderate values of
the decay rate $\Sigma$.
Periodic states  occurring  at the unstable nematic branch
are shown in Fig.~\ref{fig_per}a) and b) for three different mean
densities $\rho_0$.
The maxima of $\rho(x)$ come close
to $\rho_a$, cf. Fig.~\ref{fig_bifurc}, but
the minima of $\rho(x)$ are considerably
larger than $\rho_i$ in the case without kinetics.
The solid line in Fig.~\ref{fig_per}c) indicates the range where the  nonlinear
periodic  state is
in coexistence  with the  homogeneous  states.
For $\rho > \rho^{\ast}$, along the dashed line in Fig.~\ref{fig_per}c),
the periodic pattern becomes increasingly anharmonic,  
plateaus $\rho \sim \rho_a$ spread out and
 the valleys of low
filament density in between become less and  narrower
 by approaching the
upper end of the dashed curve   in Fig.~\ref{fig_per}c), a behavior that 
is rather uncommon \cite{CrossHo}.
At the left end of this curve the state remains periodic, the wavelength
increases and the
valleys spread out.

{\it Conclusions.-- }
A  reaction kinetically  driven
pattern forming process is predicted  near
the isotropic--nematic (I-N)  transition, which
is supported by recent experiments on actin polymerization 
\cite{Kaes:2003.101}.
A phenomenological continuum
model is introduced that reproduces
the  first order I-N   transition
in lyotropic liquid crystals, including
the phase separation in its neighborhood and being in
agreement with the statistical theory of Onsager.
Periodic solutions arise due to 
a  finite lifetime $\tau$ and a nucleation rate of the filaments.
However,  the correlation between a finite $\tau$
 and the occurrence of
periodic patterns is  independent
of the details of the model. Beyond a critical
density  $\rho_c$ the isotropic orientation
of the filaments and below a certain
 $\rho^{\ast}$ the uniform nematic state becomes unstable
against inhomogeneous perturbations. Hence for
$\rho < \rho^{\ast}$
periodic states are  favored and for our
model the wavenumber  varies as
   $q \propto \tau^{-0.17}$.
 However,
 $\rho_c$ increases
and $\rho^{\ast}$ decreases with
$\Sigma = \tau^{-1}$ 
and it may happen that
$\rho^{\ast} < \rho_c$ holds, i.e. in a certain
parameter range kinetics favors 
the uniform nematic state. Therefore, inhomogeneous states as predicted by
a linear stability analysis  of the isotropic state \cite{Doi:88.1,Liverpool:2003.2}
may be irrelevant due to nonlinear effects.
This pattern formation process
near the I-N transition is
expected to be generic and may also apply to situations
 with different transport and filament accumulation mechanisms such
as in the case of filament bundling
\cite{Takiguchi:91.1,Kruse:2001.1,Benedix:2003.101}. Even though our
description is very simplified and restricted to one spatial dimension,
 we expect that the basic physical
mechanism also plays a crucial  role for situations
with  polydisperse filament distributions
\cite{Hill:87.1,Kaes:96.1,Sollich:2002.1}, including
 living cells. Polydispersity favors periodic patterns
and together with
 higher spatial dimensions this
will give rise to an even larger variety of phenomena,
to which forthcoming works are
devoted.

We thank with great pleasure
J. Dhont, B. Gentry, M. Hammele,  J. K\"as
 and K. Kawasaki for fruitful discussions.

\bibliographystyle {prsty}

\end{document}